\documentclass[a4paper,11pt,fleqn]{article}
\usepackage{amsmath}
\usepackage{amssymb}
\usepackage{caption}
\usepackage{graphicx}
\usepackage{latexsym}
\usepackage{pslatex}
\usepackage[english]{babel}

\usepackage{graphicx} 
\usepackage{amssymb}  
\usepackage{theorem}  
\usepackage{psfrag}  
\usepackage{bm}  
\usepackage{algorithmic}
\usepackage{cite}
\usepackage{amsmath}
\usepackage{algorithm}
\usepackage{array}
\usepackage{mdwmath}
\usepackage{mdwtab}
\usepackage{eqparbox}
\usepackage{bm}   
\usepackage{subfigure}
\usepackage[dvips]{color}

\newcommand{\qed}{\nobreak \ifvmode \relax \else
      \ifdim\lastskip<1.5em \hskip-\lastskip
      \hskip1.5em plus0em minus0.5em \fi \nobreak    
      \vrule height0.5em width0.5em depth0.25em\fi}  
      
      \newcommand{\ddt}{\frac{\textnormal{d}}{\textnormal{dt}}}

\title{\LARGE \bf Predictive Control of a Permanent Magnet Synchronous Machine based on Real-Time Dynamic Optimization
}

\author{
Jean-Fran\c cois Stumper, Alexander D\"otlinger, Janos Jung and Ralph Kennel\\
Institute of Electrical Drive Systems and Power Electronics\\
Technische Universit\"at M\"unchen\\
Munich, Germany\\
Email: jean-francois.stumper@tum.de}
\date{}

\begin{document}

\setlength{\parindent}{0mm}
\baselineskip 10pt

\maketitle

\section*{Keywords}
\flqq Optimal Control\frqq, \flqq Control of Drive\frqq, \flqq Permanent Magnet Motor\frqq.

\section*{Abstract}

A predictive control scheme for a permanent-magnet synchronous machine (PMSM) is presented. It is based on a suboptimal method for computationally efficient trajectory generation based on continuous parameterization and linear programming. The torque controller optimizes a quadratic cost consisting of control error and machine losses in real-time respecting voltage and current limitations. The multivariable controller decouples the two current components and exploits cross-coupling effects in the long-range constrained predictive control strategy. The optimization results in fast and smooth torque dynamics while inherently using field-weakening to improve the power efficiency and the current dynamics in high speed operation. The performance of the scheme is demonstrated by experimental results.

\section*{Introduction}

The efforts of implementing predictive controllers in electrical drives aim at replacing the classical cascaded field-oriented control structure with PI controllers. The machine can be better exploited by improved control behavior, the system variables are optimized. In this contribution, the conventional torque and current control structure of two separate controllers is changed to multi-input multi-output (MIMO) control. By transformation into the field-oriented frame, torque generation is decoupled from flux variation, however, the current controllers are still strongly coupled, therefore a MIMO controller is advantageous. The obtained improvements are better decoupling, better current and voltage constraint handling by exploiting cross-coupling between the orthogonal components, and better power efficiency and dynamics by optimally adjusting the currents in both dynamic and steady-state operation.

The major obstacle in implementing predictive control schemes is the limited computational power, inherited by the high sampling rates. Existing implementations suffer from this restriction and can not exploit the full advantages of model predictive control (MPC). For instance, generalized predictive control (GPC) has a high optimization horizon but is unconstrained, whereas predictive torque control (PTC) is constrained but so far only reaches $2$ steps of prediction \cite{Cortes}. For good performance, both, inclusion of constraints and a high optimization horizon are required. Using control in the field-oriented frame, the analytical problem description enables using mathematical optimization algorithms. Such schemes are computationally efficient and maximize the obtained information for a given computational power. 

The online solution of the linearly constrained linear-quadratic problem, typical for MPC, requires quadratic programming (QP) algorithms, which are, however, computationally too expensive for drive systems. A recent development is the use of explicit MPC, where an offline solution is computed and stored as look-up table in the real-time controller \cite{Kuehl} \cite{inhFW}. The scheme reaches $5$ prediction steps with constraints. While several fast online algorithms were recently proposed \cite{Kouvaritakis} \cite{onlineMPC}, most of them based on gradient search algorithms, the interest in online solutions is growing. Advantages of online optimization are manipulation and adaptation of parameters, resulting in a more flexible choice of machines for the control system.

This implementation is based on a suboptimal trajectory generation algorithm presented in \cite{SK10}, embedded in a flatness-based predictive control scheme \cite{Fliess}. It is based on a continuous approach, the variables are not discretized but represented as a polynomial with undetermined coefficients. With this method, higher optimization horizons can be reached with comparably few parameters, considerably reducing the computational burden. Constraints are handled by linearization of the cost functional and the use of a linear programming (LP) solver, which is amongst the smallest and fastest numerical optimizers. As result, a (suboptimal) prediction of $2$ ms with current and voltage constraints is obtained at $8$ kHz sampling rate.


\section*{Problem Statement}


\subsection*{Machine Model}

As linear-quadratic optimization problems with linear constraints are simpler to solve in real-time, the machine model is linearized. Assuming that the rotor speed does not change too much over the optimization horizon $T$,
\begin{align}
\frac{d}{dt} \omega_M(t) &\approx 0  \;\; \Rightarrow \omega_M(t) = \textnormal{const. } \;\; \forall t\in[0,T] ,
\end{align}
the PMSM model and the voltage equations become linear. The electrical subsystem of the machine, consisting of the quadrature and direct currents $i_q$ and $i_d$ (peak values), is given as
\begin{align}
L_d \ddt i_d &=  -R i_d  + n_p \omega_M L_q i_q  + u_d   ,                 \label{eq:id} \\
L_q \ddt i_q &=  -R i_q  - n_p \omega_M L_d i_d  -n_p \omega_M K  + u_q  , \label{eq:iq} \\
      \tau_M &=   \frac{3}{2} n_p K i_q .
\end{align}
The nomenclature is shown in table II in the appendix. It is noted that the reluctance torque $\tau_M^R = \frac{3}{2} n_p (L_d-L_q)i_di_q$ is neglected, as this term is very small compared to the electromagnetic torque in surface-mounted PMSMs or in machines with small saliency. Furthermore, it would render the model nonlinear, requiring nonlinear optimization methods \cite{Delaleau}.

\subsection*{Optimization Goals and Cost Functional}

The formulation of a suitable cost functional is a key point in predictive control, as it is the only tuning possibility of the control scheme. The optimization is aiming at minimizing the control error for good dynamical performance as well as machine losses for better efficiency. Both goals are included in the cost functional. By choosing the cost functional and weights well, it is possible to find a good trade-off between both goals during transients, or eventually to fulfill both goals in steady-state. The cost functional for the predictive torque controller is
\begin{align}
 J = \int_0^T \left( P_{ctrl}(t) + W_L \cdot P_{loss}(t) \right) dt + T \cdot P_{ctrl}(T) \label{eq:J}
\end{align}
which trades off the squared control error
\begin{align}
P_{ctrl}(t) = (\tau_M-\tau_M^*)^2
\end{align}
with machine losses 
\begin{align}
P_{loss}(t) &= \frac{3}{2}R(i_d^2+i_q^2) +\frac{3}{2} n_p\omega_M k_{Fe} (\Psi_d^2+\Psi_q^2) = \frac{3}{2}R(i_d^2+i_q^2) +\frac{3}{2} n_p\omega_M k_{Fe} ((L_d i_d+K)^2+ (L_q i_q)^2)   .
\end{align}
The first term in $P_{loss}$ represents copper losses, and the second term represents the iron losses consisting of hysteresis losses. Eddy current losses are negligible on the tested machines, however, they could be included using the model presented in \cite{Eisenmodell}. Iron losses can be reduced by field-weakening, where a trade-off between copper and iron losses is found \cite{PMSMeff}. Imposing a negative direct current $i_d$, the flux magnitude in the stator is reduced while the copper losses increase. As the iron loss constant $k_{Fe}$ is not part of standard motor parameters, it has to be determined experimentally \cite{Eisenverlust}. The last term in $J$ is the end-weight of the control error and aims at reducing the steady-state control error. The weight $W_L$ was set $0.05$, the value was determined heuristically.

\begin{figure}[!ht]
  \centering
  \includegraphics[width=13cm]{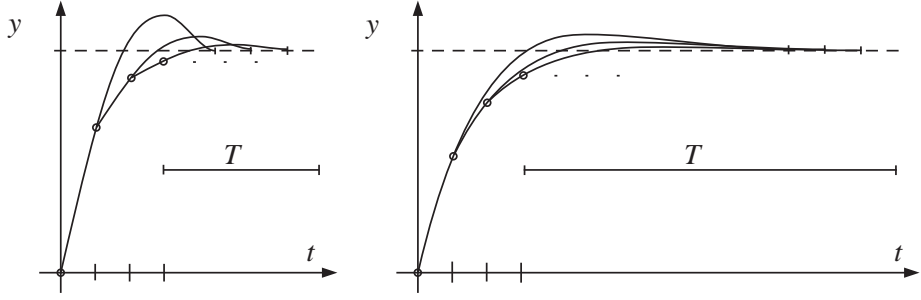}
  \caption{Open- and closed-loop trajectories in predictive control. Left: small horizon, right: high horizon.\label{fig:horizon}}
\end{figure}

The optimization horizon is set $T=2$ ms such that the cost functional includes the complete setpoint change. It is important that the optimization horizon is high enough, otherwise the open-loop and closed-loop trajectories differ and the behavior is strongly suboptimal. This is illustrated in Fig. \ref{fig:horizon}. If the horizon is too small, due to the end-weight of the control error, a significant difference between the optimized open-loop trajectories, and the closed-loop trajectories resulting from regeneration at every sampling step, appears. Then, the closed-loop trajectories simply don't fit the cost functional anymore. For a horizon higher than required for the setpoint change, the difference between open- and closed-loop trajectories becomes smaller.

\psfrag{#y}{$y$}
\psfrag{#t}{$t$}
\psfrag{#T}{$T$}

\subsection*{Current and Voltage Constraints}

The most important nonlinearities of a PMSM, in terms of control, are the  voltage and current limitations. The current constraints prevent overheating of the machine, and the voltage is limited by the maximum DC-link voltage of the inverter. The voltage constraints limit rotor speed as well as current dynamics in high-speed operation. These constraints are linearized, in order to be computationally efficiently treated. 

\begin{figure}[!ht]
  \centering
  \includegraphics[width=8.5cm]{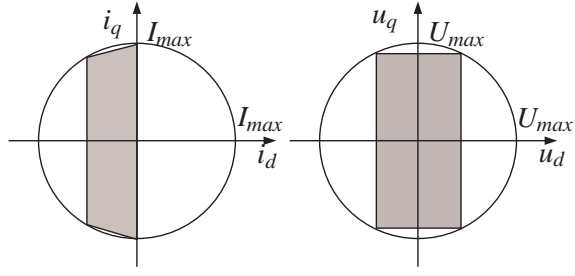}
  \caption{Linearized current and voltage constraints. Circle: feasible set of current and voltage vectors, grey: feasible set after linearization of the constraints.\label{fig:constraints}}
\end{figure}

The current range for the direct current $i_d$ is limited to $i_d^{min}\le i_d \le 0$. Only negative values of $i_d$ are desirable, as they improve power efficiency and reduce the induced voltage by weakening the flux magnitude in the stator \cite{inhFW} \cite{PMSMeff}. The lowest value $i_d^{min}$ is the optimum value at rated speed ($\frac{\partial}{\partial i_d} P_{loss}=0$) and is given as 
\begin{align}
i_d^{min} = -\frac{ L_dK }{ L_d^2 + \frac{ R }{n_p \omega_{MN} k_{Fe} } }         ,
\end{align}
which is independent of quatrature current $i_q$ as the reluctance torque was neglected. The value is doubled to enable further field-weakening to improve dynamics in high speed, an effect described in the experimental results section. For the quadrature current $i_q$, the largest possible range of values should be available. The resulting linear constraints, shown on Fig. \ref{fig:constraints}, almost completely fill the current region of interest. A linearization is thus acceptable.

The voltage linearization is a little bit more difficult. The $q$-axis should not be restricted, as the induced voltage is aligned to it and is the largest value that will appear. A steady-state analysis of the system equations (2), (3) shows that a rectangular voltage area results
\begin{align}
 R i_{d}^{min} -n_p L_q \omega_{M}^{max} i_{q}^{max}                &\le   u_d    \le   n_p L_q \omega_{M}^{max} i_{q}^{max}  ,\\
-R i_{q}^{max}  +n_pL_d\omega_{M}^{max} i_{d}^{min}  -n_p K \omega_{M}^{max}  &\le   u_q    \le   R i_{q}^{max} +n_p K \omega_{M}^{max} .
\end{align}
This rectangle is expanded such that the outer circle of the voltage limitation is hit (light grey on Fig. \ref{fig:constraints}). During dynamical transients, the voltage vector points to one of the outer corners, subsequently touching the outer limiting circle. Therefore, linearizing the voltage limits as a rectangle by the presented method, as shown on Fig. \ref{fig:constraints}, does not limit the operational range and only marginally affects dynamics. A less restrictive method is presented in \cite{onlineMPC}, where a time-varying linearization in form of a hexagon in stator frame is presented. While such a linearization is possible with the underlying predictive control algorithm, the chosen method in the ($d,q$)-frame is chosen for simplicity.

\psfrag{#iq}{$i_q$}
\psfrag{#id}{$i_d$}
\psfrag{#uq}{$u_q$}
\psfrag{#ud}{$u_d$}
\psfrag{#imx}{$I_{max}$}
\psfrag{#umx}{$U_{max}$}
\psfrag{#-il}{$-\frac{I_{max}}{2}$}

\section*{Optimal Control Algorithm}

To study real-time applicability of the presented scheme, first, the highest possible amount of optimization parameters is determined. The fastest optimizer with constraints is the widely known linear programming (LP) method. Table I shows some worst-case computational results of LP (simplex method from \cite{LPalg}) as function of the number of free parameters (CPU: $1.4$ GHz industrial PC). More parameters lead to a higher number of iterations which are also more complex; the worst-case number of iterations is the number of parameters plus the number of constraints. As in the underlying application, the constraints are decoupled, however, this worst-case is not to be expected. The maximum runtime is given by the sampling rate minus latency of input/output, therefore at $8$ kHz sampling rate, it must be less than about $110$ $\mu$s. Thus, at best, $12$ parameters can be optimized. Runtime of the predictive controller is further discussed at the end of the section.

\begin{table}[!htb]
\renewcommand{\arraystretch}{1.0}
\caption{Runtime of a linear program for some worst-case problems on a $1.4$ GHz CPU}
\label{tbl:computer}
\centering
\begin{tabular}{|c|c|c|c|}
\hline
Parameters   &  Constraints & Iterations &  Runtime [$\mu$s] \\
\hline
  20         &   48         &   67       &   769  \\
\hline
  12         &   32         &   34       &   165  \\
\hline
  8          &   14         &   10       &   35  \\
\hline
\end{tabular}
\end{table}


\subsection*{Trajectory Generation}

The trajectory generation algorithm presented in \cite{SK10}, a development related to flatness-based methods \cite{Guay}, was designed for this application. It can optimize a quadratical cost function with linear constraints. As major differences to standard algorithms, it is applying a continuous parameterization instead discretization, and the computationally efficient linear programming solver is used instead of quadratic programming or iterative gradient search.

The trajectories for the current are defined as degree $n$ power series with undetermined coefficients $\alpha_{ij}$,
\begin{align}
i_{d}(t) = \sum_{k=0}^{n} \alpha_{dk} \ \frac{t^{k}}{T^k} ,  \;\;\;\;  i_{q}(t) = \sum_{k=0}^{n} \alpha_{qk} \ \frac{t^{k}}{T^k} , \;\;\;\; t\in[0,T]  .  \label{eq:trajectory}
\end{align}
Due to the analyzed computational limitations, $n=3$ is chosen as polynomial degree. The first coefficients $\alpha_{d0}$ and $\alpha_{q0}$ are the initial conditions, and the remaining $6$ coefficients are determined by optimization. The corresponding voltages $\bm{u}_{dq}(t)$ are computed by algebraic differentiation of (\ref{eq:trajectory}) and by solving for the model equations (\ref{eq:id}) and (\ref{eq:iq}), this is also called the flatness-based approach \cite{HSR04}. This way, the voltages do not need to be represented by additional parameters.

\begin{figure}[!htb]
  \centering
  \includegraphics[width=16cm]{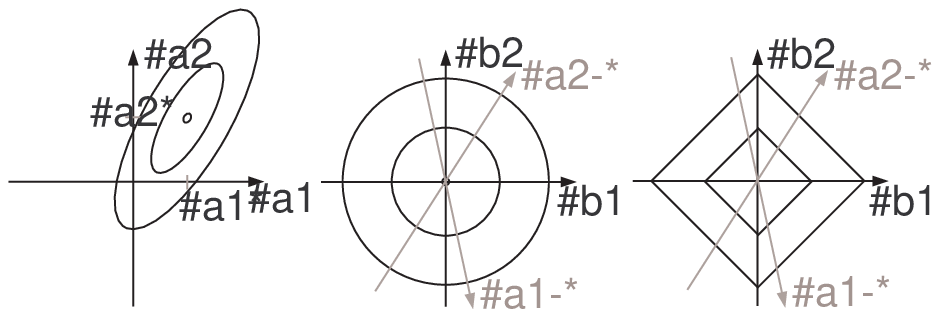}
  \caption{Trajectory generation algorithm. Left: original problem, middle: transformed least-distance problem, right: transformed and linearized problem.\label{fig:algorithm}}
\end{figure}

Substituting the variables $\bm{i}_{dq}$, $\ddt \bm{i}_{dq}$ and $\bm{u}_{dq}$ by the found functionals in the cost functional $J$ (\ref{eq:J}) and constraints can easily be done with a computer-algebra tool (for instance Maplesoft's$^\text{\textregistered}$ Maple$^{\textnormal{TM}}$). The cost functional $J$ is then a quadratic function of the unknown parameters, of the machine parameters, the measured currents, the speed $\omega_M$ and the torque reference $\tau_M^*$. Graphically, $J$ can be represented as in Fig. \ref{fig:algorithm} (left).

As $J$ is convex, the unconstrained optimum is found algebraically by solving first-order necessary conditions, in this case, the solution of a system of linear equations by matrix inversion. Then, by an affine coordinate transformation, the problem can be reformulated as least-distance problem, i.e. a quadratical cost describing the distance to the unconstrained optimum. As result, the cost functional looks much simpler, see Fig. \ref{fig:algorithm} (middle).

\psfrag{#a1}{$\alpha_1$}
\psfrag{#a2}{$\alpha_2$}
\psfrag{#b1}{$\beta_1$}
\psfrag{#b2}{$\beta_2$}
\psfrag{#a1*}{\textcolor[gray]{0.5}{$\alpha_1^*$} }
\psfrag{#a2*}{\textcolor[gray]{0.5}{$\alpha_2^*$} }
\psfrag{#a1-*}{\textcolor[gray]{0.5}{$\alpha_1$-$\alpha_1^*$} }
\psfrag{#a2-*}{\textcolor[gray]{0.5}{$\alpha_2$-$\alpha_2^*$} }

In the next step, the least-distance problem is linearized around the unconstrained optimum, see Fig. \ref{fig:algorithm} (right). As the coordinate transformation to the least-distance problem is an affine transformation, and as the linearization doesn't affect the constraints, it is obvious that the linear constraints remain linear after this transformation. One point is, however, more difficult, the parameterization of the constraints with the polynomial coefficients, presented in \cite{SK10}. The linearization of the cost function inherits a large error in the value of $J$, but the values of the coefficients $\alpha$ are not affected that much: the least-distance problem is not so much different in the linear form as it would have been in quadratical form. Furthermore, see that a difference only appears if a constraint is active, the unconstrained optimum is the same. It can be shown that the resulting cost inherited by the linearization is
\begin{align}
J' = J_0 + 2 n  \cdot J_C
\end{align}
in the worst case, where $J_0$ is the unconstrained cost, and $J_C$ the extra cost when considering constraints. The suboptimality is therefore bounded. As $n=3$, there are $6$ parameters in the quadratical cost function. During the linearization, this number is increased to $12$, as in the LP standard form, only positive values are possible. After the transformation, the problem is reformulated in standard form for linear programming, and a simplex solver \cite{LPalg} can be run. To visualize the suboptimality, a comparison (simulations) between the linearized and the original problem is presented in \cite{SK10}.

All presented computations are done using Maplesoft's$^\text{\textregistered}$ Maple$^{\textnormal{TM}}$, from where the matrices for the LP solver are generated using a C code generation toolbox. The real-time software thus consists of a simplex tableau assignment, which is automatically generated code, a simplex LP solver and some post-processing. As the assignment is based on symbolic calculations, the machine parameters can be changed online.

\subsection*{Predictive Control}

\begin{figure}[!ht]
  \centering
  \includegraphics[width=15cm]{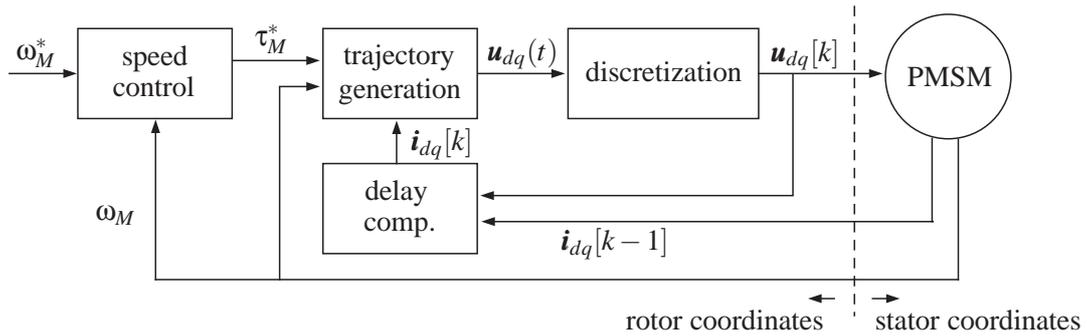}
  \caption{Control structure of the predictive torque controller cascaded by PI speed control.\label{fig:structure}}
\end{figure}

The trajectory generation scheme is embedded in a predictive controller. The control structure is shown in Fig. \ref{fig:structure}. A cascaded control structure is chosen as speed is assumed constant for trajectory generation. As the mechanical plant is generally only roughly known, this structure is advantageous. Model-based control can be used for the machine as the parameters are known, but for the mechanical part, any robust feedback controller can be chosen.

\psfrag{#speed}{speed}
\psfrag{#ctrl}{control}
\psfrag{#traj}{trajectory}
\psfrag{#gen}{generation}
\psfrag{#delay}{delay}
\psfrag{#comp}{comp.}
\psfrag{#disc}{discretization}
\psfrag{#PMSM}{PMSM}

\psfrag{#fieldframe}{rotor coordinates}
\psfrag{#stator frame}{stator coordinates}

\psfrag{#w}{$\omega_M$}
\psfrag{#w*}{$\omega_M^*$}
\psfrag{#t*}{$\tau_M^*$}
\psfrag{#u(t)}{$\bm{u}_{dq}(t)$}
\psfrag{#u}{$\bm{u}_{dq}[k]$}
\psfrag{#i[k]}{$\bm{i}_{dq}[k]$}
\psfrag{#i}{$\bm{i}_{dq}[k-1]$}

First, measurement and control timing is analyzed. Two timing sequences are shown on Fig. \ref{fig:results} (a) and (b). The interrupt-based control system triggers an interrupt every $125\mu$s (green signal). At this instant, the applied voltage command (magenta signal) is modulated by space-vector modulation. At the same instant, the A/D conversion of the current measurements (blue signal) is performed to avoid the impact of current ripples. After the interrupt, the current $i_q[k]$ is available and the controller (cyan signal) is started to compute the next voltage command $u_q[k+1]$, to be modulated at the next interrupt. The \textit{computational delay} is accounted with a delay compensation technique that is applied to generate $i_q[k+1]$ as feedback value for the control law to compute $u_q[k+1]$ \cite{Moon}, it corresponds to a prediction of one sampling step with the model, the current $i_q[k]$ as well as the previously commanded voltage $u[k]$. Furthermore, see that the response to the commanded voltage $u_q[k]$ is $i_q[k+1]$ which is available one interrupt later, this is the \textit{plant delay} and it is naturally included by recalculating the trajectory at every sampling step, see Fig. \ref{fig:horizon}. From the predicted trajectory, the applied voltage is $\bm{u}_{dq}[k] = \bm{u}_{dq}(0)$.

From the timing sequences, interesting insight into the computational demands of the algorithm is gained. The first part of the cyan signal shows the calculation time for the simplex tableau initialization, it takes about $10\mu$s. Included in these calculations, which are results of symbolical calculations, is a calculation of the unconstrained optimum and the linearization of the problem. The second and biggest part of the cyan signal is the runtime of the linear program. At the beginning of Fig. \ref{fig:results} (a), where voltage and current are both zero, it is only about $20\mu$s, but to calculate the voltage step at $2000$ rpm shown on Fig. \ref{fig:results} (b), more iterations are involved as many constraints are active, and the computation time rises to almost $60\mu$s. The total time of the interrupt handling, the simplex initialization, the LP solver and the post-processing sum up to almost $100\mu$s in the worst case, therefore up to $80\%$ of the available time is used.

\section*{Results}

A surface-mounted PMSM with parameters shown in table II is used. It is coupled to a load drive such that an arbitrary load torque or speed can be applied. The algorithm is implemented on a PC-104 based real-time system with a $1.4$ GHz CPU described in \cite{Nael}. The voltage limitation was set to $75\%$ of the possible $330$ V to clearly see behavior of saturation in control, accordingly, rated speed is reduced from $3000$ to $2250$ rpm.

\begin{figure*}[!ht]
\centering
\subfigure[Timing at $0$ rpm.]{\includegraphics[width=7.8cm,height=5.5cm]{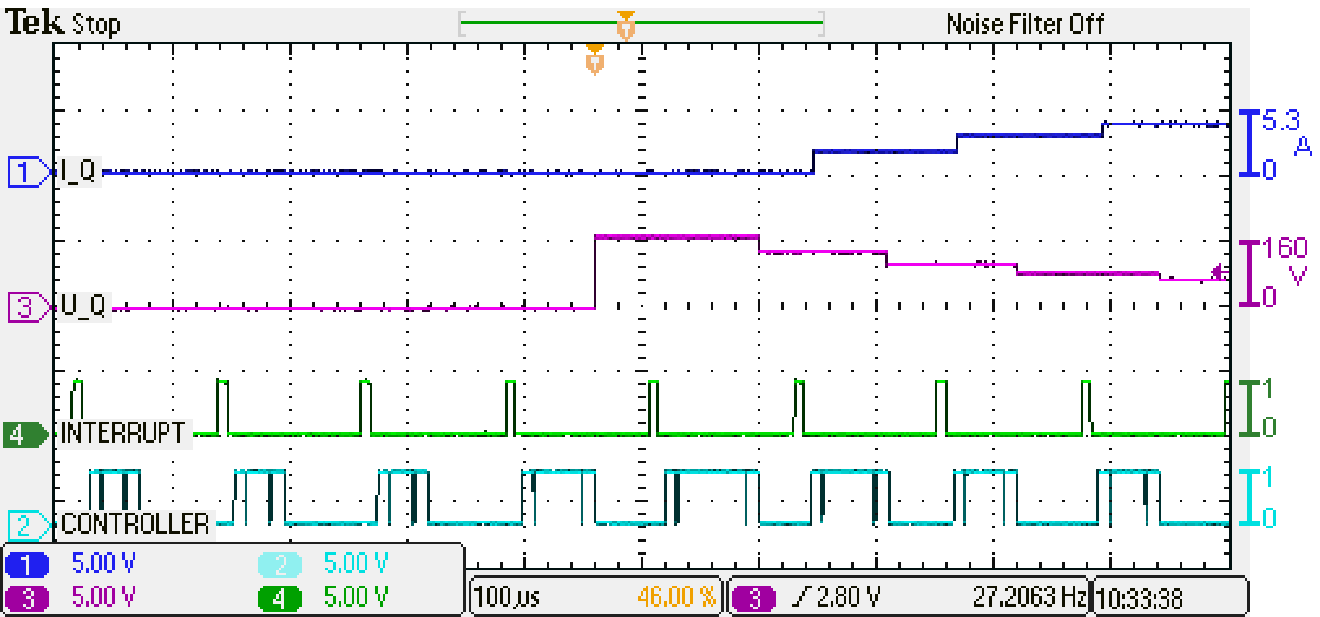}}
\subfigure[Timing at $2000$ rpm.]{\includegraphics[width=7.8cm,height=5.5cm]{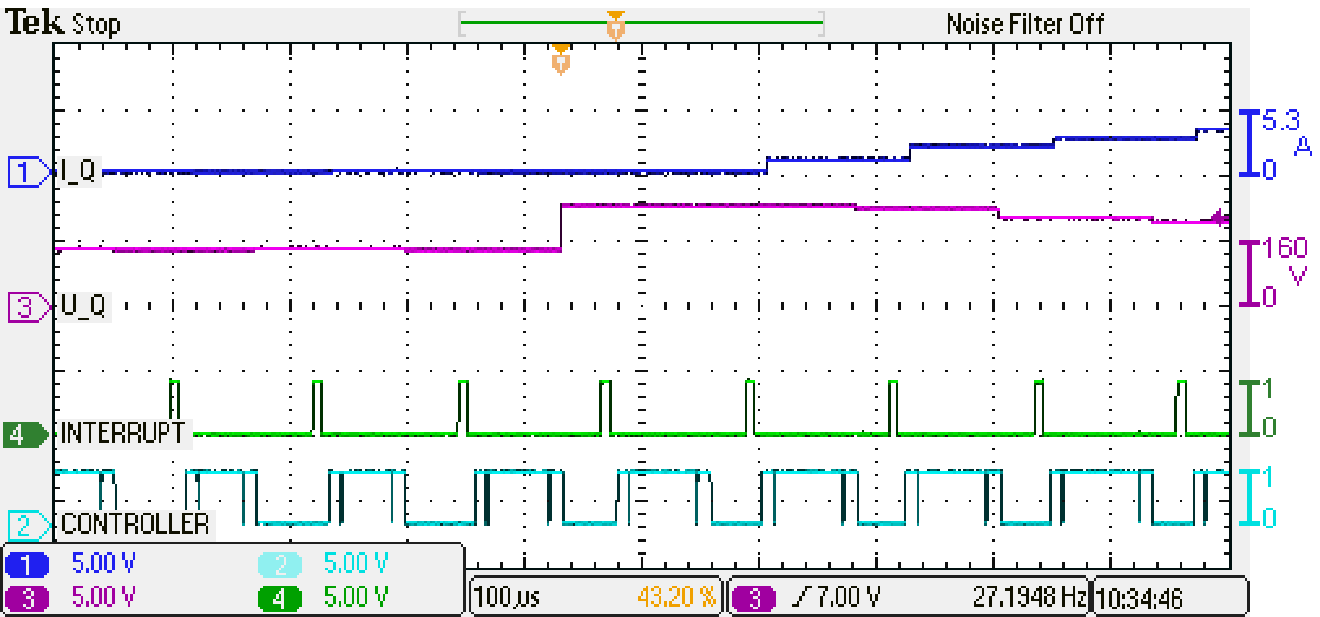}}
\subfigure[Speed steps from $0$ to $1000$ to $2000$ and $0$ rpm.]{\includegraphics[width=7.8cm,height=5.5cm]{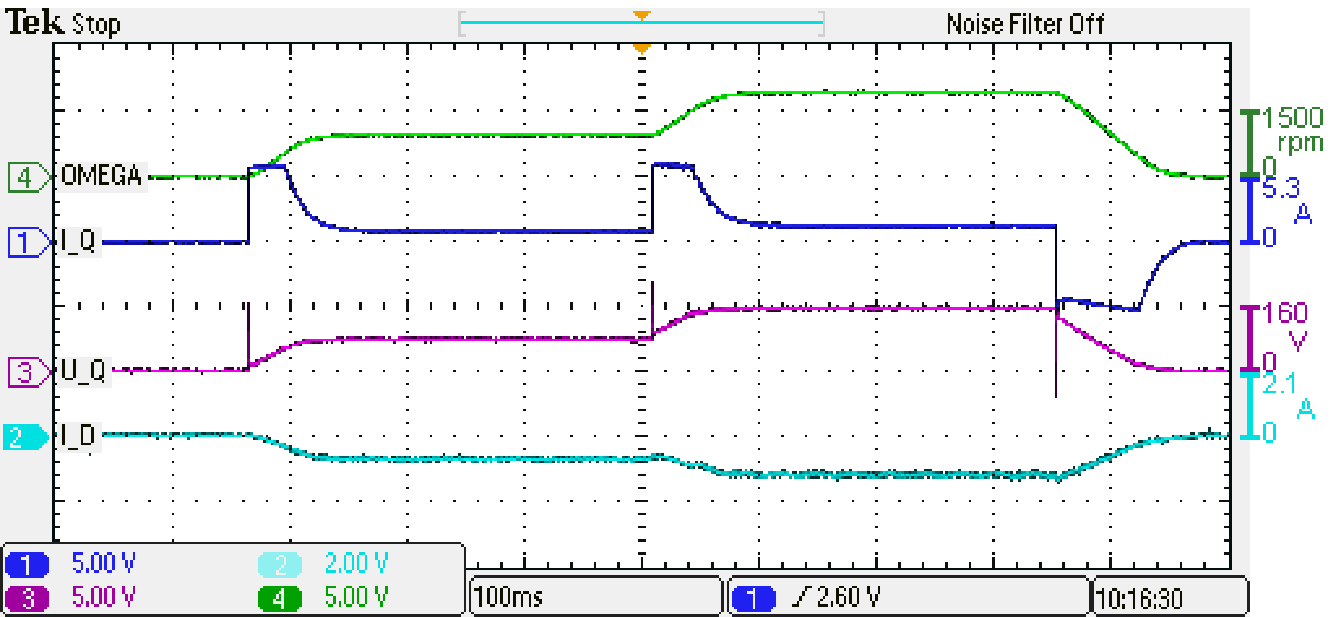}}
\subfigure[Torque step at zero speed.]{\includegraphics[width=7.8cm,height=5.5cm]{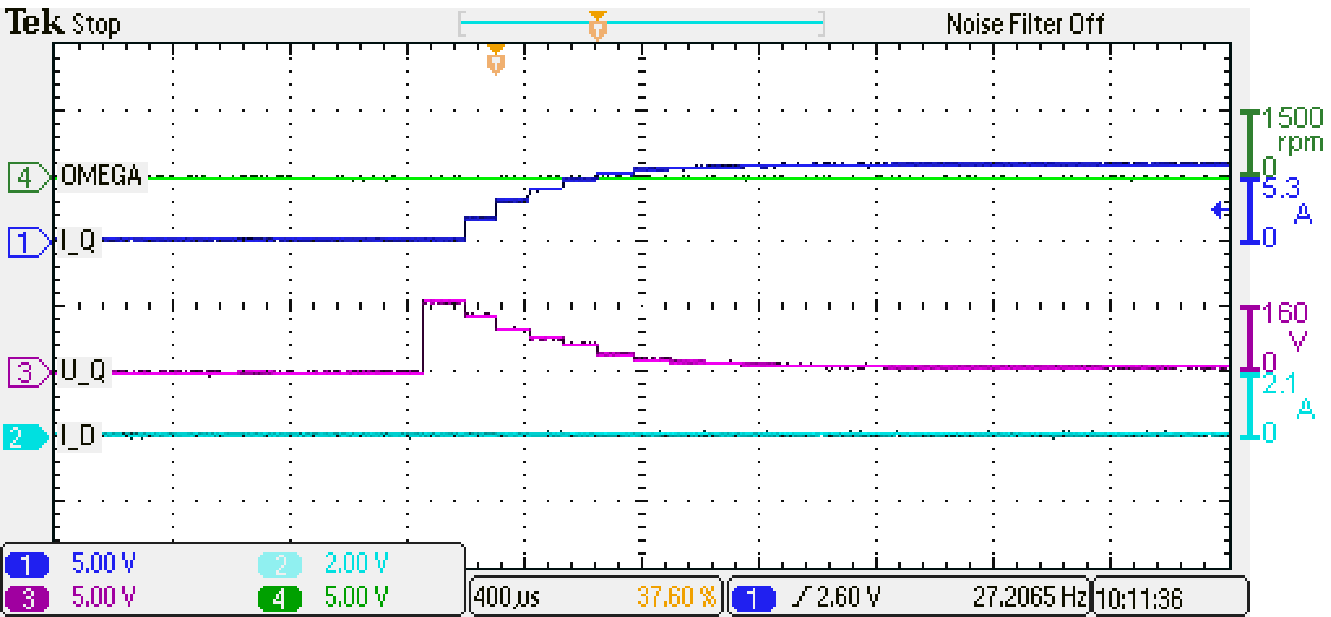}}
\subfigure[Torque step at $2000$ rpm.]{\includegraphics[width=7.8cm,height=5.5cm]{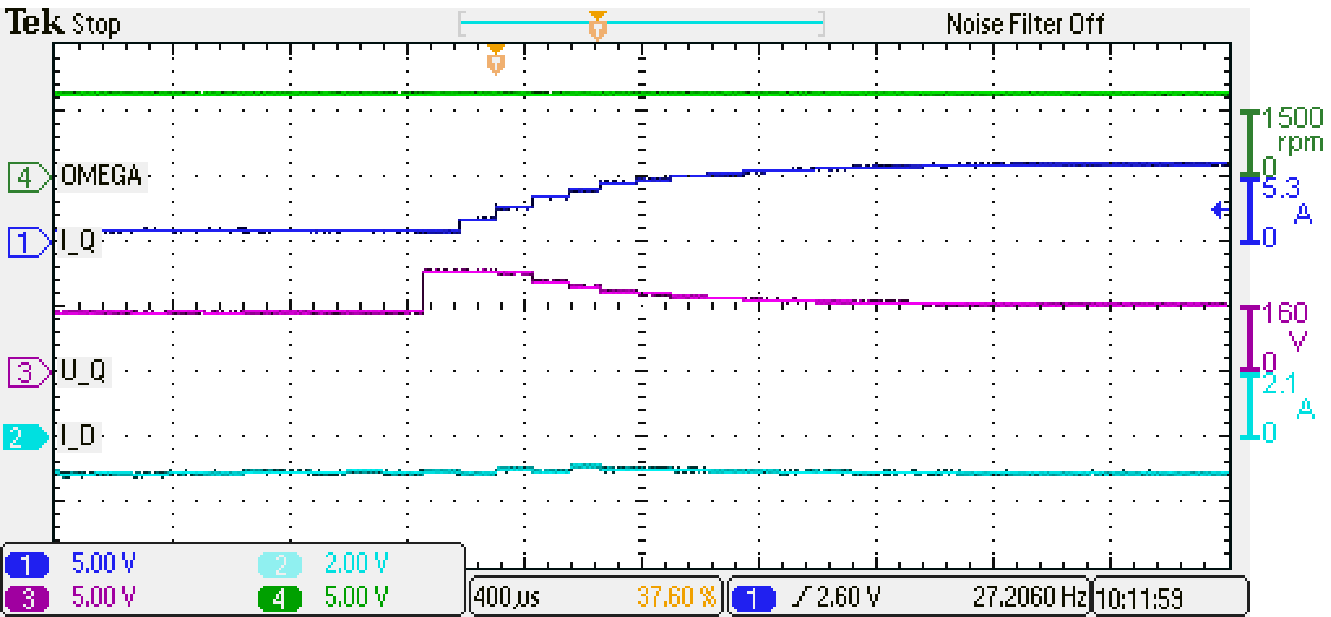}}
\subfigure[Torque step at $2400$ rpm.]{\includegraphics[width=7.8cm,height=5.5cm]{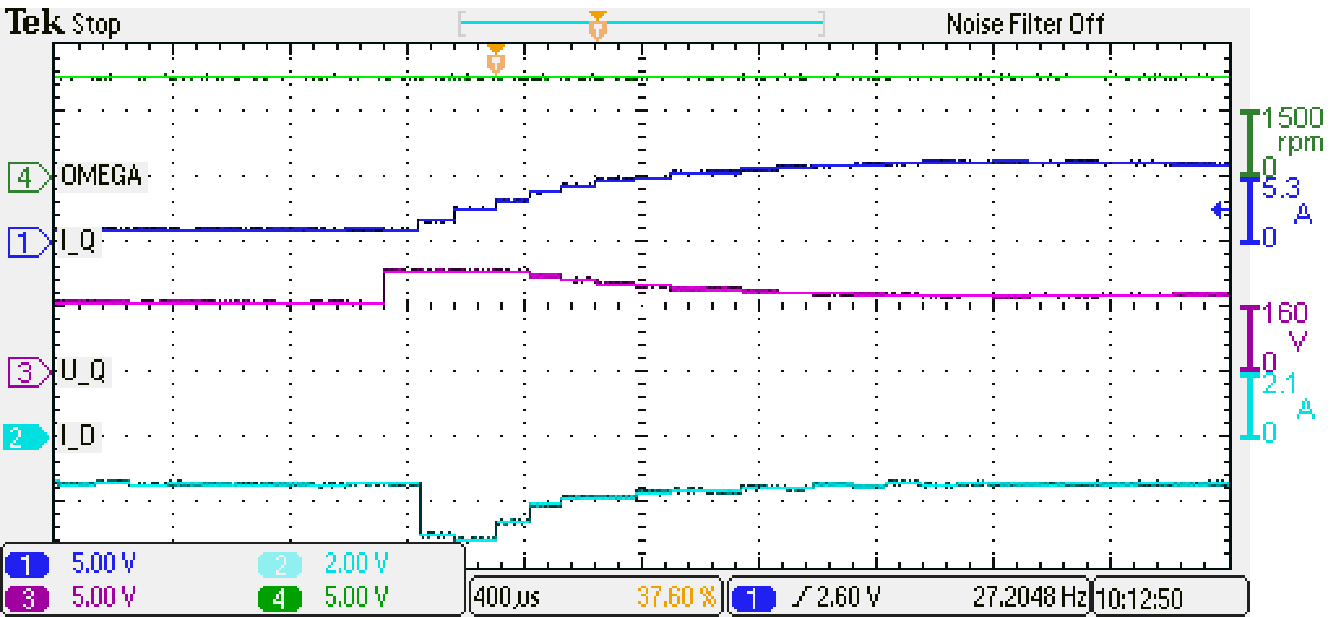}}
\caption{Experimental results of the predictive control scheme. Subfigures (a)-(b): blue: quadrature current $i_q$ without delay compensation ($5.3$ A/div), cyan: quadrature voltage $u_q$ (160 V/div), green: interrupt handling, cyan: control law computation. Subfigures (c)-(f): green: rotor speed $\omega_M$ ($1500$ rpm/div), blue: quadrature current $i_q$ with compensation ($5.3$ A/div), cyan: quadrature voltage $u_q$ (160 V/div), cyan: direct current $i_d$ ($2.1$ A/div).\label{fig:results}}
\end{figure*}

Experimental results of the proposed scheme are shown in Fig. \ref{fig:results}. Subfigure (c) shows the response to two subsequent speed reference steps, the load drive is deactivated. The torque is increased rapidly by a high but feasible voltage peak. Interestingly, no overshoot arises, even though the speed change is performed quickly. The direct current is proportional to the speed and thereby reduces iron losses which are considerable at high speeds. Losses are decreased by about $4\%$, and the efficiency is improved by about $0.5\%$ at $2000$ rpm. Better results are obtained on machines with higher inductances \cite{PMSMeff}.

The next three subfigures (d), (e) and (f) show fast torque transients at zero, medium and high speed, respectively. The PMSM is in torque control mode while the load drive keeps speed constant at $0$, $2000$ and $2400$ rpm, respectively. The current components are well decoupled, a fast current change on the quadrature axis does not affect the direct axis at all in (d) and (e). With two separate PI controllers, a short current excursion would be seen on the direct axis during the torque transient. Again, the current on the direct axis $i_d$ is proportional to the speed. Furthermore, the torque is fast and at the same time smooth, the voltage becomes proportionally small for smaller control errors -- a nice characteristic of quadratic cost functionals, compared to linear cost functions which result in deadbeat behavior and are more sensitive to uncertainties \cite{Moon}. 

On subfigures (d) and (e), behavior with active voltage constraint is the same as when using standard saturation or anti-windup strategies. On subfigure (f), however, a different behavior is seen, the direct current $i_d$ is reduced to perform field-weakening. This implies that the stator induced voltage is reduced on the quadrature axis, see eq. (\ref{eq:iq}). Thereby the derivative of quadrature current $\ddt i_q$ is higher and the torque-generation dynamics increased, at the cost of higher copper losses on the direct axis. Without additional field-weakening, the reference torque would not be reached after the optimization horizon of $2$ ms, thereby the end-weight in $J$ oversizes the loss term. Therefore, in this predictive control implementation, field-weakening not only improves efficiency, but also improves dynamics by exploiting cross-coupling between the orthogonal current components to optimally bypass the voltage saturation. 

It is also possible to operate the PMSM beyond rated speed by steadily doing field-weakening to bypass the voltage saturation on the quadrature axis, as shown in \cite{inhFW}. It is remarked that the current on the direct axis $i_d$ has no reference, its value is obtained from the optimization of the cost function. Therefore the method works well and is numerically stable; the optimal value follows inherently.

\section*{Conclusion}
A predictive control scheme for a PMSM was introduced. Based on suboptimal real-time optimization, the currents and voltages are computed according to a cost functional. The prediction horizon is $2$ ms at a sampling rate of $8$ kHz, and voltage and current constraints are respected. The advantages of the long-range constrained predictive MIMO control scheme can be concluded as follows: improved decoupling and accounting of cross-coupling, precise measurement and control timing, respecting current and voltage constraints, fast and smooth dynamical behavior, improved power efficiency by field weakening, and improved dynamics close to voltage saturation by field-weakening.

Furthermore, it was shown that it is possible to implement long-range MPC with constrained online-optimization even on fast-sampling systems such as electrical drives.

\section*{Acknowledgments}
This work was supported through the National Research Fund of Luxembourg under grant PhD-08-070.

\clearpage


\section*{Appendix: Machine Parameters}

\begin{table}[!h]
\renewcommand{\arraystretch}{1.0}
\caption{Nominal Parameters of the Synchronous Machine\label{tbl:SMparameter}}
\centering
\begin{tabular}{|l||c|}
\hline
Manufacturer $\&$ Model  & Merkes MT5 1050   \\
\hline
Rated Power $P_N$         &  $2640$ W  \\ 
\hline
Rated Torque  $\tau_{MN}$ &  $8.4$ Nm  \\
\hline
Rated Current   (peak)      &  $5.6$ A \\
\hline
Rated Speed $\omega_{MN}$ &  $3000$ rpm \\
\hline
Pole Pairs  $n_p$         &  $3$ \\
\hline
Rated Voltage $U_{N}$   (peak)  &  $560$ V \\
\hline
Stator Inductance $L_d$, $L_q$     &  $4.8$, $7.2$ mH \\
\hline
Stator Resistance $R$     &  $0.92$ $\Omega$ \\
\hline
Motor Constant $K$  (peak)   &  $0.334$ Vs \\
\hline
Iron  Loss Constant (Hysteresis) $k_{Fe}$        &  $1.27$ $\frac{\textnormal{A}}{\textnormal{Vs}}$ \\
\hline
\end{tabular}
\end{table}

\end{document}